\documentclass[%
 reprint,
nofootinbib,
 amsmath,amssymb,
 aps,
]{revtex4-2}

\usepackage{graphicx}
\usepackage[caption=false]{subfig}
\usepackage{dcolumn}
\usepackage{bm}
\usepackage{hyperref}

\usepackage{xcolor}
\usepackage{multirow}

\begin{document}

\title{On the reliability of parameter estimates in the first observing run of Advanced LIGO}

\author{Suman Kulkarni}%
\email{suman.kulkarni@students.iiserpune.ac.in}
\affiliation{%
Indian Institute of Science Education and Research, Homi  Bhabha road, Pashan, Pune 411008, India}%
\author{Collin D. Capano}%
\email{collin.capano@aei.mpg.de}
\affiliation{%
Max-Planck-Institut f{\"u}r Gravitationsphysik (Albert-Einstein-Institut), D-30167 Hannover, Germany\\
Leibniz Universit{\"a}t Hannover, D-30167 Hannover, Germany}%

\date{\today}

\begin{abstract}

Accurate parameter estimation is key to maximizing the scientific impact of gravitational-wave astronomy. Parameters of a binary merger are typically estimated using Bayesian inference. It is necessary to make several assumptions when doing so, one of which is that the detectors output stationary Gaussian noise. We test the validity of these assumptions by performing percentile-percentile tests in both simulated Gaussian noise and real detector data in the first observing run of Advanced LIGO (O1). We add simulated signals to 512s of data centered on each of the three events detected in O1 --- GW150914, GW151012, and GW151226 --- and check that the recovered credible intervals match statistical expectations. We find that we are able to recover unbiased parameter estimates in the real detector data, indicating that the assumption of Gaussian noise does not adversely effect parameter estimates. However, we also find that both the parallel-tempered sampler \texttt{emcee\_pt} and the nested sampler \texttt{dynesty} struggle to produced unbiased parameter estimates for GW151226-like signals, even in simulated Gaussian noise. The \texttt{emcee\_pt} sampler does produce unbiased estimates for GW150914-like signals. This highlights the importance of performing percentile-percentile tests in different targeted areas of parameter space.

\end{abstract}

\maketitle

\section{\label{sec:intro}Introduction}

To date, the LIGO \cite{TheLIGOScientific:2014jea} and Virgo \cite{TheVirgo:2014hva} observatories have detected over 50 gravitational waves from binary black hole and binary neutron star mergers \cite{LIGOScientific:2018mvr,Nitz:2018imz,Venumadhav:2019lyq,Nitz:2019hdf,Abbott:2020niy}. These detections have opened a new window on to the Universe, offering insights that would not be possible from electromagnetic observations alone. This is because gravitational waves carry information about the source binaries' parameters. By careful observation of a gravitational wave's frequency and amplitude evolution, one can estimate a binary's masses, spins, location, and other properties. From these measurements it is possible to infer the distribution of black holes in the universe \cite{Abbott:2020gyp}, measure the equation of state of dense nuclear matter \cite{De:2018uhw,Abbott:2018exr,Capano:2019eae}, constrain cosmological parameters \cite{Schutz:1986,Abbott:2017xzu,Abbott:2019yzh}, and (by allowing modifications to model waveforms) test general relativity (GR) in the strong-field regime \cite{Abbott:2020jks,LIGOScientific:2019fpa,Abbott:2018lct,TheLIGOScientific:2016src}. Accurate parameter estimation is therefore critical to the success of gravitational-wave astronomy.

Bayesian inference is the standard method by which physical parameters are extracted from gravitational waves. Given some observed data and a gravitational-wave model, a posterior probability distribution is obtained on the parameters describing the binary. This posterior is then marginalized to obtain credible intervals on specific parameters. Stochastic samplers are needed to fully map out the posterior distribution due to the high dimensionality and complicated topology of the parameter space. Several assumptions are made in this process, such as: the template gravitational-wave model is an accurate representation of the signal; the digital output from the detectors is accurately calibrated to the strain induced by the passing gravitational wave; the stochastic sampler produces an accurate representation of the posterior.

One of the most critical assumptions made when estimating parameters is that, in the absence of a signal, the detectors output wide-sense stationary Gaussian noise. This leads to a canonical likelihood function that can be evaluated numerically, providing an algorithm for estimating posterior distributions. However, the LIGO and Virgo detectors are known to produce a number of non-Gaussian noise transients (``glitches'') \cite{Nuttall:2015dqa,TheLIGOScientific:2016zmo,TheLIGOScientific:2017lwt,TheLIGOScientific:2017qsa,Cabero:2019orq}. Indeed, search pipelines must employ a number of bespoke statistics in order to overcome these glitches \cite{Allen:2004gu,Nitz:2017svb,Usman:2015kfa,sachdev2019gstlal}. Non-Gaussian transients are generally less of a concern for parameter estimation. Since gravitational waves from binary mergers have finite duration within the LIGO/Virgo sensitive frequency band, estimating the parameters of a binary typically involves between O(10s) and O(100s) of data. At current sensitivity, both the signal rate and glitch rate is low enough that it is rare for a non-Gaussian transient to occur during the time of interest for a parameter estimation analysis. Even so, it has happened.

A large transient occurred in the Livingston detector $\sim1\,$s before the merger of GW170817~\cite{TheLIGOScientific:2017qsa}. This nearly caused the signal to be missed by low-latency search pipelines~\cite{GBM:2017lvd}. For the followup parameter estimation analyses, it was necessary to remove the transient from the data by fitting a glitch model to the transient~\cite{TheLIGOScientific:2017qsa,Cornish:2014kda}. Without this removal procedure the inferred parameters of the binary would have been biased \cite{Pankow:2018qpo}. This, in turn, would have yielded misleading answers to key questions, such as what the equation of state of dense nuclear matter is.

The transient that occurred during GW170817 was loud enough that it could be identified and removed. However, its presence raises the question, what if there is less-obvious non-Gaussian noise present in the data containing a binary merger? Such noise, being unmodelled, could adversely affect the reported credible intervals. This would be particularly problematic for tests of GR using gravitational waves.

Several previous studies have tested the validity of Bayesian inference tools used for gravitational-wave parameter estimation~\cite{Sidery:2013zua,Veitch:2014wba,Berry:2015pe_color,Biwer:2018osg,Smith:2019ucc,Romero-Shaw:2020owr}. The standard test is to perform a percentile-percentile test. This involves generating a number of simulated signals, which are drawn from some prior distribution. These are added to some data; a Bayesian inference analysis is then performed on each signal separately, using the same prior. This yields credible intervals on each of the signals' parameters. If all of the assumptions made in the analysis are valid, then $X\%$ of the signals should fall within the $X$-percentile credible interval for any given parameter.

The primary focus of previous studies employing percentile-percentile tests was to validate software tools used for gravitational-wave Bayesian inference~\cite{Sidery:2013zua,Veitch:2014wba,Biwer:2018osg,Smith:2019ucc,Romero-Shaw:2020owr}. This was accomplished by adding simulated signals to simulated Gaussian noise, colored by power spectral densities (PSDs) representative of the LIGO and Virgo detectors. A violation of a percentile-percentile test under these conditions would indicate that some aspect of the software --- e.g., the stochastic sampler used --- was not properly recovering the posterior distribution. A percentile-percentile test was also used in Ref.~\cite{Berry:2015pe_color} to study the accuracy of parameter estimates of signals that were anticipated to be detected in Advanced LIGO given non-Gaussian noise. As Advanced LIGO had not begun yet, that study used Initial LIGO data that was recolored to resemble the expected power spectral density of Advanced LIGO.

In this paper we, for the first time, perform percentile-percentile tests in real Advanced LIGO data. Our aim is to test the assumption that the detectors' output can be modelled as a gravitational-wave plus stationary Gaussian noise in data surrounding identified gravitational-wave events. We do this by performing identical percentile-percentile tests in real and simulated Gaussian noise. Should the test be violated in the former and not the latter, it would indicate that the detector data is not sufficiently Gaussian during the observation time. We apply this test to the three gravitational-wave events that were detected during the first observing run (O1) of Advanced LIGO: GW150914~\cite{Abbott:2016blz}, GW151012~\cite{TheLIGOScientific:2016pea,Nitz:2018imz,LIGOScientific:2018mvr}, and GW151226~\cite{Abbott:2016nmj}.

Our paper is structured as follows: in Section~\ref{sec:bayesianinference} we review Bayesian inference and its application to gravitational waves. In Section~\ref{sec:methods} we detail the methods used in this study. The results of our study are discussed in Section~\ref{sec:results}. Finally, in Section~\ref{sec:outlook}, we discuss the implications of our results and prospects for future studies.

\section{\label{sec:bayesianinference}Parameter Estimation using Bayesian Inference}

Consider a network of $N$ detectors labelled with indices i = 1,..., $N$. The data collected at the $i^{th}$ detector is a time series comprising of a signal under some waveform model $H$ along with the detector noise:
\begin{equation*}
    d_{i}(t) = n_{i}(t)+s_{i}(t).
\end{equation*}
Here, $n_{i}(t)$ is the noise observed at the $i^{th}$ detector and $s_{i}(t)$ is the gravitational waveform obtained under the model used. We denote the collection of data at all detectors by $\mathcal{D}(t)$ and the set of parameters for the waveform model by $\textbf{v}$.

Applying the Bayes' Theorem, the posterior probability density function is
\begin{equation}
    \underbrace{P(\textbf{v}\vert\mathcal{D}(t),H)}_\text{Posterior} = \frac{\overbrace{P(\mathcal{D}(t)\vert\textbf{v},H)}^\text{Likelihood}\overbrace{P(\textbf{v}\vert H)}^\text{Prior}}{\underbrace{P(\mathcal{D}(t)\vert H)}_\text{Evidence}}.
    \label{bayes}
\end{equation}
The prior indicates our knowledge of the parameters in a given model before analysing the data. Assuming circular orbits, models describing binary black hole mergers involve 15 parameters: the component masses $m_{1,2}$, the magnitude and orientation of the component spins, the luminosity distance $d_{L}$, the right ascension $\alpha$, the declination $\delta$, the polarization $\psi$, the binary inclination angle $\iota$, the coalescence time $t_{c}$, and the phase at the time of coalescence $\phi$.

We use similar priors as in Ref.~\cite{Nitz:2019hdf}. For BBH mergers it is common to use uniform priors on the component source masses, choosing the bounds such that all regions with non-zero posterior support are within the boundaries. For the magnitudes of each component spin vector, we use uniform priors on $a_{1,2}$ $\in [0.0, 0.99)$. For the other two components of each spin vector, we use a uniform solid angle prior, which assumes a uniform distribution for the azimuthal angle $\theta^{azimuthal}_{1,2} \in [0, 2\pi]$ and a sine-angle distribution for the polar angle $\theta^{polar}_{1,2}$. We make use of a uniform sky location prior, which assumes a uniform distribution on $\alpha \in [0,2\pi)$ and a cosine-angle distribution for $\delta$. The polarization angle $\psi$ is uniform $\in [0, 2\pi)$ and the inclination $\iota$ uses a sine-angle prior. Uniform priors are used for the coalescence time $\pm$0.1 s around the time of the merger. A uniform prior on $[0, 2\pi)$ is used for $\phi$ and analytically marginalized over, as discussed later in this section.

For the luminosity distance, a uniform prior on comoving volume was used in Ref.~\cite{Nitz:2019hdf} (which was converted to luminosity distance by assuming a standard $\Lambda$CDM cosmology \cite{Ade:2015xua}), with bounds chosen to enclose the posterior for each event. However, in this work, we use a prior uniform in the $\log_{10}$ of the comoving volume. We do this so as to sample higher SNRs; see Section \ref{credible} for details. 

We assume that each detector outputs independent, wide-sense stationary Gaussian noise in the absence of a signal. Under this assumption the likelihood function is
\begin{align}
    P(&\mathcal{D}(t)|\textbf{v},H) \propto \nonumber \\
    & \exp\left[ -\frac{1}{2}\sum_{i=1}^{N} \langle \tilde{d}_{i}(f) - \tilde{s}_{i}(f,\textbf{v}),\,\tilde{d}_{i}(f) - \tilde{s}_{i}(f,\textbf{v})\rangle \right]
\label{eqn:likelihood}
\end{align}
The inner product $\langle \tilde{a}_i(f),\,\tilde{b}_i(f)\rangle$ is defined as
\begin{equation*}
    \langle\tilde{a}_i(f),\,\tilde{b}_i(f)\rangle = 4 \Large{\Re\int_{0}^{\infty} \frac{\tilde{a}_i^{*}(f)\tilde{b}_i(f)}{S^{(i)}_{n}(f)} df} 
\end{equation*}
where $S^{(i)}_{n}(f)$ is the PSD of the noise in the $i^{th}$ detector.

We use the median-mean variation of Welch's method as described in Ref.~\cite{Allen:2005fk} to estimate the PSD in each detector. As in Ref.~\cite{Nitz:2019hdf}, $\pm256\,$s of data centered on each simulated signal is broken up into overlapping $8\,$s segments for this purpose. The O1 LIGO detectors' PSD grows very rapidly at frequencies below $\sim20\,$Hz. Consequently, we use a lower-frequency cutoff of $20\,$Hz when generating template waveforms $\tilde{s}(f)$ and evaluating inner products.

As was done in Ref.~\cite{Nitz:2019hdf}, we use the IMRPhenomPv2 waveform model~\cite{Hannam:2013oca,Schmidt:2014iyl} to generate template waveforms when evaluating the likelihood. We also use this model to generate our simulated signals. IMRPhenomPv2 models the inspiral, merger, and ringdown of the dominant gravitational-wave mode emitted by circular, precessing binary black holes. Due to some simplifying assumptions made in the model, the waveform's dependence on the coalescence phase $\phi$ can be written as
\begin{equation}
\label{eqn:simple_phase}
    \tilde{s}_{i}(\textbf{v},f,\phi) = \tilde{s}_{i}^{0}(\textbf{v},f,\phi=0)e^{i\phi}.
\end{equation}
It is possible to analytically marginalize over the phase for signals of this form. Substituting Eq.~\eqref{eqn:simple_phase} into the likelihood Eq.~\eqref{eqn:likelihood} and marginalizing the posterior over $\phi$ using a uniform prior yields
\begin{multline}
    \log P(\textbf{v}|\mathcal{D})\propto \log P(\textbf{v}) +
    \log I_0\left[\left|\sum_i O(s^0_i, d_i)\right|\right] \\
     -\frac{1}{2}\sum_i\left[ \left<s^0_i,\, s^0_i\right> +
                            \left<d_i,\, d_i\right> \right],
\end{multline}
where \begin{equation*}
   O(s^0_i, d_i) \equiv 4 \int_0^\infty
    \frac{\tilde{s}_i^*(f; \textbf{v},\phi=0)\tilde{d}_i(f)}{S^{(i)}_n(f)}\mathrm{d}f,
\end{equation*}
and $I_0$ is the modified Bessel function of the first kind. We use this form of the likelihood function in our analysis, obviating the need to sample over phase. This substantially reduces computational cost, as the phase is a difficult parameter for stochastic samplers to measure.

\section{\label{sec:methods}Methods}
\subsection{Obtaining Samples}\label{samples}

We use the PyCBC Inference software library to perform our analysis~\cite{Biwer:2018osg}. PyCBC Inference provides a collection of tools for performing Bayesian inference on gravitational waves, as well as doing percentile-percentile tests on simulated signals. It has support for multiple stochastic samplers. In this analysis, we use the parallel-tempered, ensemble Markov-chain Monte Carlo (MCMC) sampler \verb+emcee_pt+ \cite{ForemanMackey:2012ig,Vousden:2015}, with 200 walkers and 20 temperatures.

To ensure that the samples we obtain for the posterior are independent of their initial position of the chains, we use the \verb+max_posterior+ and \verb+n_acl+ burn-in tests. The \verb+max_posterior+ test requires that all chains sample at least one point with a prior-weighted log likelihood within $N_{D}/{2}$ of the maximum over all chains, where $N_{D}$ is the number of dimensions. The \verb+n_acl+ test ensures that the length of each chain is greater than 10 times the autocorrelation length, as calculated using the second half of the chain. If so, samples in the second half of the chain are retained. The sampler is considered burned-in at the first iteration that passes both tests. Post burn-in samples are thinned by their autocorrelation length so that we only use independent samples to estimate the posterior probability density function. We run the sampler until we obtain 1400 -- 2000 independent samples.

\subsection{Credible intervals and the percentile-percentile test}\label{credible}

Once the posterior is estimated, we obtain a credible interval on each parameter by marginalizing over all of the other parameters. The $X\%$ credible interval gives the region of parameter space that contains $X\%$ of the marginalized posterior probability of that parameter. In other words, we expect the true value of the parameter to be within the $X\%$ credible interval with $X\%$ probability. Hence, the credible intervals provide a useful way to test whether a parameter estimation analysis is biased. 

We perform a Bayesian analysis on a set of simulated events. For each parameter, we plot the fraction of signals in which the true parameter value lies in a credible interval as a function of credible intervals. This is called a percentile-percentile plot. If the recovery is unbiased, we expect the plot for each parameter to follow the $y=x$ line, with some fluctuations due to noise. To quantify the deviations from the expected $y=x$ line, we perform a Kolmogorov-Smirnov (KS) test. This gives us a two-tailed p-value for each parameter.

The p-value gives the probability of obtaining a percentile-percentile curve at least as extreme as the observed curve under the null hypothesis that the analysis provides an unbiased estimate of the parameter. A very small p-value (lower than a level of significance $\alpha$) points to an outcome which is very unlikely under the null hypothesis, allowing us to reject the null hypothesis with a maximum Type I error of $\alpha$. We apply this test to each parameter, yielding 13 p-values for each event. If the analysis is unbiased, we in turn expect these p-values to be uniformly distributed between $0$ and $1$. We therefore perform another KS-test on the set of p-values to obtain a single ``p-value of p-values", by which we can evaluate the probability that the analysis is unbiased. This method was used in Ref.~\cite{Biwer:2018osg} to verify that \verb+emcee_pt+ provides an unbiased estimate of gravitational-wave parameters in Gaussian noise, using a prior similar to that used for GW150914.

For each event (GW150914, GW151012 and GW151226), we generate 100 simulated signals (``injections'') with parameters drawn from the same mass and spin priors used for that event in Ref.~\cite{Nitz:2019hdf}. The coalescence times of the injections are drawn from a prior uniform in an interval of $\pm256\,$s centered on the reported coalescence time of each event. We choose this time window because it was the amount of time used to estimate the PSD for each event in the original analysis~\cite{2-OGC}. As was done in the original analyses, we use a uniform prior on the coalescence time $t_c \in t_0 + [-0.1, 0.1)\,$s, where $t_0$ is GPS time of the simulated signal. Since the prior is centered on signals' injected time, we do not include $t_c$ in the percentile-percentile tests.

\begin{figure}
    \centering
    \includegraphics[scale=0.55]{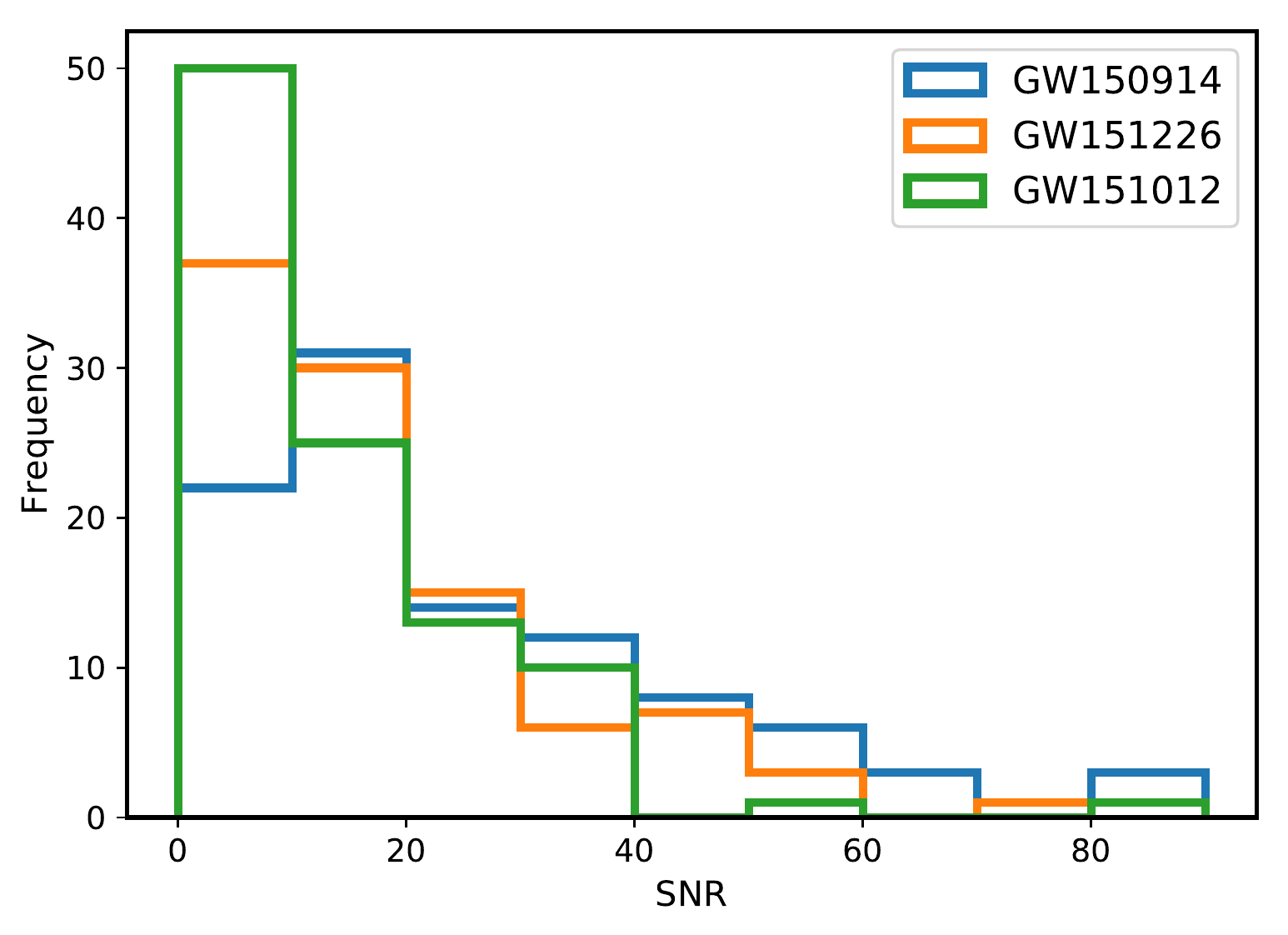}
    \caption{Distribution of the ``optimal'' SNRs of the injections used in our analysis. Here, SNR is defined as $\sqrt{\sum_i \left<\tilde{s}_i,\,\tilde{s}_i\right>}$, where the sum is over the number of detectors and $s_i$ is the simulated waveform. The bounds on the comoving volume prior are chosen so that majority of the injections have an SNR close to that of the actual event. }
    \label{fig:SNR}
\end{figure}

The original analysis used a prior uniform in comoving volume to encapsulate the posterior~\cite{Nitz:2019hdf}. Instead, we use a prior uniform in $\log_{10}$ of the comoving volume. This is done to avoid having too many injections with low SNR. We choose the bounds such that the majority of the injections have SNRs representative of their actual event while ensuring that the 90\% credible interval of the reported distances are contained in the bounds. The distribution of the SNRs of the resulting injections is shown in Fig.~\ref{fig:SNR}. The prior bounds used on the source masses and the comoving volume (expressed in terms of the luminosity distance) are uniform over the intervals listed in Table~\ref{tab:priors}. For the remaining parameters, we use the priors discussed in Section \ref{sec:bayesianinference}. 

The amount of time analyzed for each event needs to be large enough to encapsulate the longest-duration waveform allowed by the prior. The in-band waveform duration of binary mergers is approximately inversely proportional to the chirp mass of the binary. For GW150914 and GW151012, we use an analysis duration of 8 seconds around each simulated signal, which is sufficiently long given our low frequency cutoff of $20\,$Hz. GW151226, being lower mass, requires a longer analysis time. To limit the amount of time that needed to be analyzed, a constraint on the detector-frame chirp mass $\mathcal{M}^{\rm det} = (m^{\rm det}_1 m^{\rm det}_2)^{3/5}/(m^{\rm det}_{1} + m^{\rm det}_{2})^{1/5}$ was applied to the GW151226 analysis in both Ref.~\cite{Nitz:2019hdf} and in the original publication by the LIGO and Virgo collaborations~\cite{Abbott:2016nmj}. We include the same constraint --- $\mathcal{M}^{det} \in [8.7, 10.7]\,\mathrm{M}_\odot$ --- as used in Ref.~\cite{Nitz:2019hdf} here~\cite{2-OGC}. So as not to bias the percentile-percentile test, this constraint is applied both when analyzing each simulated signal, and when drawing the parameters for the signals. With this constraint in this place, we need only to analyze 12 seconds of data around each signal in the GW151226 percentile-percentile test.

We set up two types of runs for each event --- one in stationary Gaussian noise and one in real detector noise. In the former, we generate stationary Gaussian noise colored by the PSDs representative of the sensitivity of the LIGO detectors at the time of detection of the respective event. We then add the same simulated signals to the two noise runs and perform a parameter estimation analysis on each signal. We obtain credible intervals on all the parameters and construct the percentile-percentile plot. We also find the p-values of p-values as described in Section~\ref{credible}. We checked that the presence of the original events in the real detector noise runs do not significantly affect the p-values and their interpretation.

When analyzing each simulated signal, we re-estimate the PSDs using a window of $\pm256\,$s centered on that signal, as described above. Consequently, our test makes use of up to $\pm512\,$s of real data centered on the original events. Data is downloaded from the Gravitational-wave Open Science Center (GWOSC)~\cite{Abbott:2019ebz}.

\setlength{\tabcolsep}{6pt}
\renewcommand{\arraystretch}{1.2}

\begin{table}
    \centering
    \begin{tabular}{|c|c|c|c|}
         \hline
         Parameter & GW150914 & GW151226 & GW151012 \\
         \hline
         $m_1~(\mathrm{M}_\odot)$ & [10, 80]& [7, 50] & [10, 80]  \\
         $m_2~(\mathrm{M}_\odot)$ & [10, 80]& [3, 15] & [10, 80] \\
         \multirow{2}{7em}{$V_{C}$ converted to $d_{L}$ (Mpc)} & \multirow{2}{*}{[50,700]} & \multirow{2}{*}{[150, 700]} & \multirow{2}{*}{[300, 1500]}\\
         & & & \\
         \hline
    \end{tabular}
    \caption{Source mass and comoving volume priors used in our analysis. The comoving volume is expressed in terms of the luminosity distance.}
    \label{tab:priors}
\end{table}

\subsection{Test on simulated glitches}\label{glitches}

As a proof of principle, we perform a percentile-percentile test on simulated non-Gaussian noise, which we create by adding glitches to Gaussian noise. We use the same realization of Gaussian noise and the same injections as used for the GW150914 analysis. Glitches are created by using the BayesWave~\cite{Cornish:2014kda} reconstruction of the glitch that occurred in the Livingston detector during GW170817~\cite{TheLIGOScientific:2017qsa, Pankow:2018qpo, gw170817glitch}. We add the transient at random times to our simulated Hanford and Livingston data, with an average rate of one glitch per 16 seconds in each detector. Each glitch is given a random phase offset that is drawn uniformly in $[0, 2\pi)$, and we randomly scale the amplitude of each transient by using a uniform prior in $[0, 1]$ for the scale factor. The glitch times, phase, and amplitude are uncorrelated between the two detectors.

\begin{figure}
    \includegraphics[width=\columnwidth]{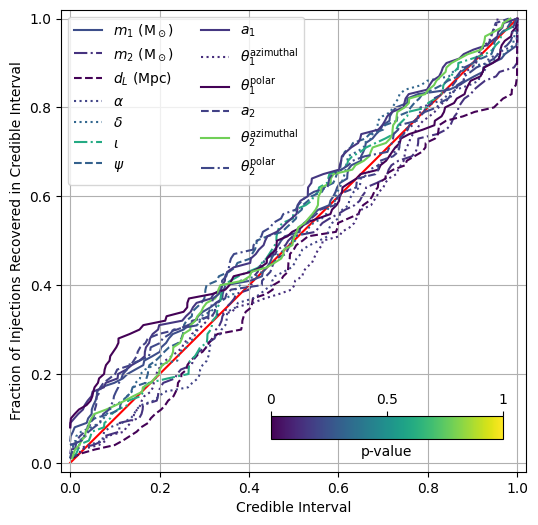}
    \caption{Percentile-percentile plot for simulated non-Gaussian noise. Shown are the fraction of simulated signals with parameter values recovered in a credible interval as a function of credible intervals for each parameter. If the parameters were recovered exactly, the plot would follow the line \emph{y=x} as indicated by the diagonal line. Simulated glitches were added to the same Gaussian noise as used in Fig~\ref{fig:GW150914}, and the same simulated signals were used to perform the test. We perform a KS-test to obtain a p-value for each parameter, which is indicated by the color bar. We then perform a KS-test on the set of p-values to obtain a p-value of p-values of 0.001, indicating that the noise fails the percentile-percentile test, as expected.}
    \label{fig:gw150914_glitchy}
\end{figure}

The percentile-percentile plot for the simulated glitch data is shown in Fig.~\ref{fig:gw150914_glitchy}. We obtain a p-value of p-values for this analysis of 0.001. Compared to the results using Gaussian noise --- see Fig.~\ref{fig:GW150914} and the first column of Table~\ref{gauss_table} --- the non-Gaussian noise is clearly failing the percentile-percentile test, as expected.

\section{\label{sec:results}Results}

\begin{figure*}[h]
    \centering
    \subfloat[Gaussian Noise\label{sfig:150914A}]{%
        \includegraphics[scale=0.45]{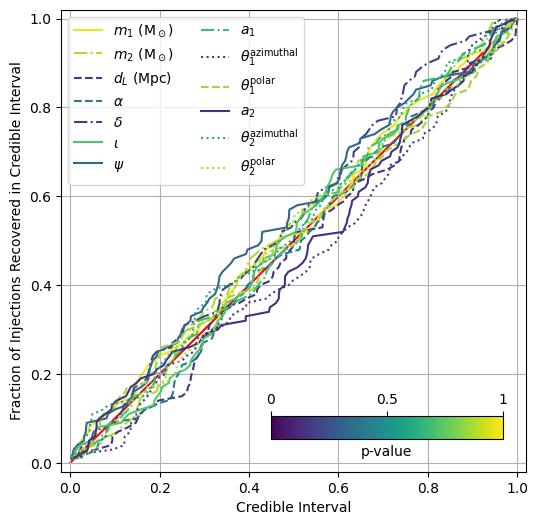}
    }\hfill
    \subfloat[Real Noise\label{sfig:150914B}]{%
        \includegraphics[scale=0.45]{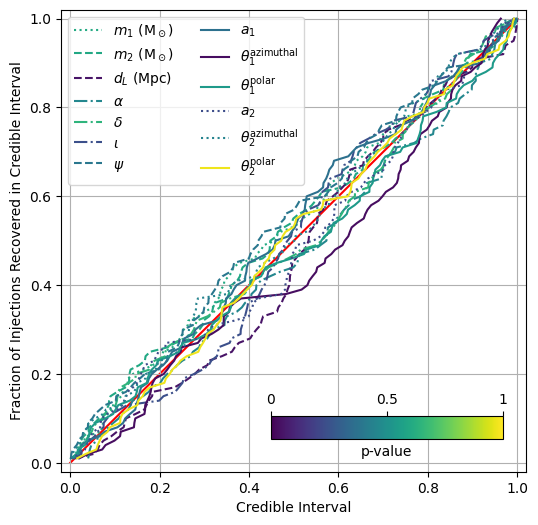}
    }
    \caption{\textbf{GW150914:} Plot of the fraction of simulated signals with parameter values recovered in a credible interval as a function of credible intervals for each parameter. If the parameters were recovered exactly, the plot would follow the line \emph{y=x} as indicated by the diagonal line. For each parameter, a KS test is performed between the recovered curve and the diagonal line to obtain a two-tailed p-value, which is indicated by the color bar.}
    \label{fig:GW150914}
\end{figure*}

\begin{figure*}[h]
    \centering
    \subfloat[Gaussian Noise]{%
        \includegraphics[scale=0.45]{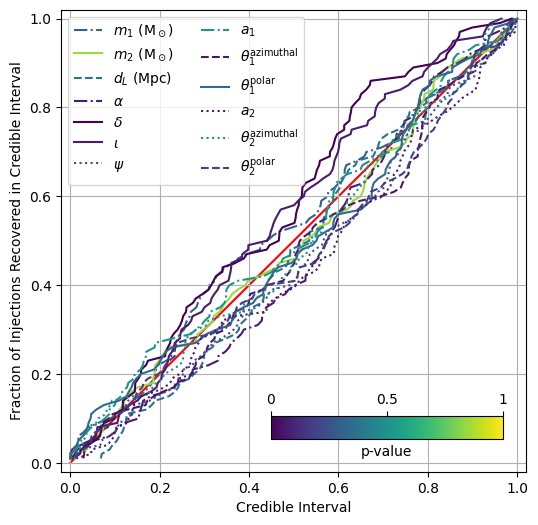}
    }\hfill
    \subfloat[Real Noise\label{sfig:151226B}]{%
        \includegraphics[scale=0.45]{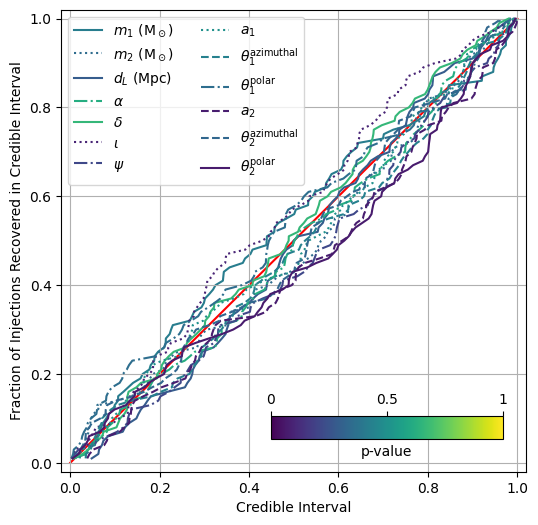}
    }
    \caption{\textbf{GW151226:} Plot of the fraction of simulated signals with parameter values recovered in a credible interval as a function of credible intervals for each parameter. If the parameters were recovered exactly, the plot would follow the line \emph{y=x} as indicated by the diagonal line. For each parameter, a KS test is performed between the recovered curve and the diagonal line to obtain a two-tailed p-value, which is indicated by the color bar.}
    \label{fig:GW151226}
\end{figure*}

The results are summarized in Tables~\ref{gauss_table} and \ref{real_table} for the Gaussian noise and real noise runs, respectively. Percentile-percentile plots for GW150914 are shown in Fig.~\ref{fig:GW150914} and for GW151226 in Fig.~\ref{fig:GW151226}. The percentile-percentile plot for GW151012 (not shown) is qualitatively similar to GW150914.

For GW150914 and GW151012, we find a p-value of p-values of 0.86 and 0.98 in Gaussian noise, respectively. We therefore find no reason to reject the null hypothesis that the sampler provides an unbiased estimate of the parameters. In real noise, we obtain a p-value of p-values of 0.25 and 0.94 for the two events, respectively. Although the p-value of p-values is lower for GW150914 in real noise, it is not small enough to reject the null hypothesis that the data containing GW150914 is stationary and Gaussian.

The results from GW151226 are less clear, however.  The p-value of p-values is less than 0.1 in both Gaussian and real noise. This indicates that the \verb+emcee_pt+ sampler is not fully converging on the posterior under this prior. Given the low p-values from \verb+emcee_pt+, we decided to try the \verb+dynesty+ nested sampler \cite{speagle:2019} with the GW151226 prior in Gaussian noise.\footnote{Due to the way \texttt{dynesty} samples the parameter space it was necessary to remove the constraint on chirp mass when testing it. This meant increasing the analyzed time to 30 seconds, and generating a new set of simulated signals.} Several previous studies have shown that \texttt{dynesty} can produce unbiased parameter estimates of gravitational wave sources~\cite{Smith:2019ucc,Romero-Shaw:2020owr,Finstad:2020sok}. However, in this case, we find that \verb+dynesty+ yields even worse results than \texttt{emcee\_pt}, with a p-value of p-values of 0.001. We did not try to analyze real noise with \verb+dynesty+ as a result.

\begin{figure*}[h]
    \centering
    \subfloat[Gaussian Noise]{%
        \includegraphics[scale=0.45]{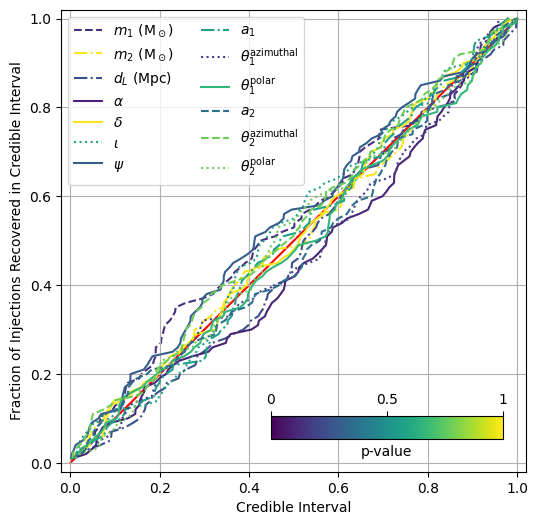}
    }\hfill
    \subfloat[Real Noise\label{sfig:151012B}]{%
        \includegraphics[scale=0.45]{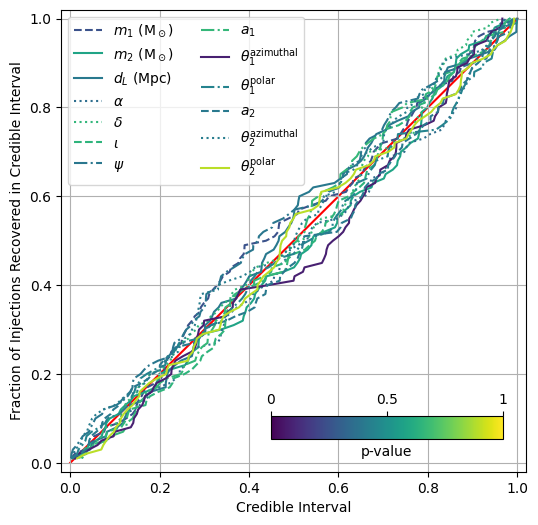}
    }
    \caption{\textbf{GW150914 Shifted:} Plot of the fraction of simulated signals with parameter values recovered in a credible interval as a function of credible intervals for each parameter. If the parameters were recovered exactly, the plot would follow the line \emph{y=x} as indicated by the diagonal line. For each parameter, a KS test is performed between the recovered curve and the diagonal line to obtain a two-tailed p-value, which is indicated by the color bar.}
    \label{fig:GW150914_shifted}
\end{figure*}

The low-frequency noise in the Livingston detector had larger amplitude during GW151226 than it did during GW150914. In order to test whether the small p-values for GW151226 are due to the different PSD at those times, we shifted the coalescence times of the injections used in the GW150914 runs to those used in the GW151226 runs. We then perform the percentile-percentile test in Gaussian noise using the same noise realizations as was used to analyze the GW151226-like injections, and with the \verb+emcee_pt+ sampler. The resulting percentile-percentile plot is shown in Fig.~\ref{fig:GW150914_shifted}; p-values are reported in the last column of Table~\ref{gauss_table}. Doing this, we obtained a p-value of p-values of 0.92. The low p-values for the GW151226-like signals is therefore not due to the changing PSD.

Since the GW150914 injections shifted to the GW151226 times passed the percentile-percentile test in Gaussian noise, we repeat the analysis in the real noise around GW151226, again using the \verb+emcee_pt+ sampler. The results are reported in Fig.~\ref{fig:GW150914_shifted} and the last column of Table~\ref{real_table}. We obtain a p-value of p-values of 0.31 in this case. We therefore cannot rule out the null hypothesis that the real noise around GW151226 is stationary and Gaussian. 

\section{Summary and Outlook}
\label{sec:outlook}

We have performed percentile-percentile tests in both simulated and real noise using the same priors that were used to analyze GW150914, GW151012, and GW151226. Simulated signals were added to a $512\,$s block of time centered on each event. Comparing percentile-percentile test results from real data to simulated noise, we find no reason to reject the null hypothesis that the detector data was sufficiently stationary and Gaussian at these times to produce unbiased parameter estimates for these events.

We find that both the \verb+emcee_pt+ sampler and the \verb+dynesty+ nested sampler struggle to produce unbiased parameter estimates for signals similar to that of GW151226. Previous studies have shown that both of these samplers pass percentile-percentile tests for GW150914-like signals~\cite{Biwer:2018osg,Smith:2019ucc,Romero-Shaw:2020owr}. In addition, Ref.~\cite{Finstad:2020sok} recently showed that the \verb+dynesty+ sampler passes percentile-percentile tests for binary neutron star and neutron-star--black-hole binaries. This suggests that the difficulty with GW151226-like signals is not due to their low mass, or even large mass ratio (Ref.~\cite{Finstad:2020sok} allowed mass ratios up to 10:1). One major difference from our work is that Ref.~\cite{Finstad:2020sok} only considered aligned-spin signals. That reduced the number of parameters involved in their analysis as compared to ours. In addition, spin precession adds more structure to waveforms, leading to a more complicated likelihood topology. This is particularly true of lower-mass and larger mass-ratio signals, which is targeted by the GW151226 prior. However, while our results are suggestive, determining if precession is the primary difficulty for these samplers will require more study.

Regardless of the cause, the poor percentile-percentile test results we obtain for GW151226-like signals with both \verb+emcee_pt+ and \verb+dynesty+ highlights the need for better stochastic samplers. Even if a sampler is shown to produce unbiased parameter estimates for some region of parameter space, as both these samplers have, it does not mean that the sampler will do so for all parts of parameter space. For this reason, the gravitational-wave community should strive to continually perform these tests as new waveform models and more sensitive detectors become available. The primary hurdle to performing percentile-percentile tests is the computational cost involved. However, new methods for fast likelihood estimation~\cite{Zackay:2018qdy}, and the ease with which newer inference toolkits~\cite{Biwer:2018osg,Romero-Shaw:2020owr} can parallelize over many cores make regular percentile-percentile tests more feasible.

We emphasize that our results in real data do not \emph{prove} that the detectors' noise is stationary and Gaussian, only that we have no reason to doubt that they are. It is possible that a non-Gaussian noise component exists in the data during the inspiral and merger of one these events that is simply missed by our analysis, or is not detected due to the statistical nature of our test. Even so, our results give confidence that assuming stationary Gaussian noise has not lead to biased parameter estimates in O1. We plan to extend this test to other detected events in the future.

\section*{Acknowledgements}

We thank Sumit Kumar, Alexander Nitz, and Badri Krishnan for useful suggestions and insights. This work was made possible by the summer intern research program at the AEI Hannover. S.K. would like to acknowledge the funding provided by DAAD-WISE. We are grateful to the computing team at the AEI Hannover for maintaining the Atlas computer cluster, which was used to carry out all analyses. This research has made use of data obtained from the Gravitational Wave Open Science Center (https://www.gw-openscience.org), a service of LIGO Laboratory, the LIGO Scientific Collaboration and the Virgo Collaboration. LIGO Laboratory and Advanced LIGO are funded by the United States National Science Foundation (NSF) as well as the Science and Technology Facilities Council (STFC) of the United Kingdom, the Max-Planck-Society (MPS), and the State of Niedersachsen/Germany for support of the construction of Advanced LIGO and construction and operation of the GEO600 detector. Additional support for Advanced LIGO was provided by the Australian Research Council. Virgo is funded, through the European Gravitational Observatory (EGO), by the French Centre National de Recherche Scientifique (CNRS), the Italian Istituto Nazionale della Fisica Nucleare (INFN) and the Dutch Nikhef, with contributions by institutions from Belgium, Germany, Greece, Hungary, Ireland, Japan, Monaco, Poland, Portugal, Spain.

\bibliography{paper}

\begin{table*}
 \begin{center}
 \begin{tabular}{ |c|c|c|c|c| } 
 \hline
 Parameter & GW150914 & GW151226 & GW151012 & GW150914 prior \\
 & prior & prior & prior & $t_{c}$ shifted \\
 \hline
 $m_1~(\mathrm{M}_\odot)$ & 0.975 & 0.314 & 0.814 & 0.137 \\
 $m_2~(\mathrm{M}_\odot)$ & 0.876 & 0.850 & 0.483 & 0.990 \\
 $d_L$ (Mpc) & 0.154 & 0.377 & 0.032 & 0.248 \\
 $\alpha$ & 0.405 & 0.076 & 0.701 & 0.103 \\
 $\delta$ & 0.192 & 0.001 & 0.342 & 0.990 \\
 $\iota$ & 0.731 & 0.076 & 0.122 & 0.567 \\
 $\psi$ & 0.326 & 0.238 & 0.268 & 0.296 \\
 $a_{1}$ & 0.693 & 0.518 & 0.419 & 0.567 \\
 $\theta_1^\mathrm{azimuthal}$ & 0.176 & 0.064 & 0.693 & 0.176 \\
 $\theta_1^\mathrm{polar}$ & 0.850 & 0.333 & 0.966 & 0.659 \\
 $a_{2}$ & 0.134 & 0.065 & 0.829 & 0.359 \\
 $\theta_2^\mathrm{azimuthal}$ & 0.609 & 0.487 & 0.021 & 0.775 \\
 $\theta_2^\mathrm{polar}$ & 0.873 & 0.184 & 0.526 & 0.775 \\
 \hline
 p-values of p-values & 0.855 & 0.019 & 0.978 & 0.915 \\
 \hline
 \end{tabular}
 \caption{P-values obtained for each parameter when recovered from Gaussian noise colored by the power-spectral densities representative of the detectors at the time of the respective events. The last column corresponds to results from the injections used in the GW150914-like run shifted to the coalescence times used in the GW151226-like injections. We perform a KS-test comparing the set of p-values obtained for each event to a uniform distribution and report the p-value of p-values in the last row.}
 \label{gauss_table}
 \end{center}
 \end{table*}

 \begin{table*}
 \begin{center}
 \begin{tabular}{ |c|c|c|c|c| } 
 \hline
 Parameter & GW150914 & GW151226 & GW151012 & GW150914 prior \\
 & prior & prior & prior & $t_{c}$ shifted \\
 \hline
 $m_1~(\mathrm{M}_\odot)$ & 0.609 & 0.420 & 0.901 & 0.253 \\
 $m_2~(\mathrm{M}_\odot)$ & 0.600 & 0.351 & 0.722 & 0.584 \\
 $d_L$ (Mpc) & 0.054 & 0.290 & 0.282 & 0.405 \\
 $\alpha$ & 0.464 & 0.619 & 0.169 & 0.357 \\
 $\delta$ & 0.651 & 0.668 & 0.487 & 0.659 \\
 $\iota$ & 0.243 & 0.101 & 0.051 & 0.651 \\
 $\psi$ & 0.398 & 0.219 & 0.276 & 0.398 \\
 $a_{1}$ & 0.371 & 0.501 & 0.973 & 0.668 \\
 $\theta_1^\mathrm{azimuthal}$ & 0.035 & 0.434 & 0.494 & 0.089 \\
 $\theta_1^\mathrm{polar}$ & 0.542 & 0.356 & 0.676 & 0.449 \\
 $a_{2}$ & 0.233 & 0.078 & 0.802 & 0.405 \\
 $\theta_2^\mathrm{azimuthal}$ & 0.449 & 0.332 & 0.034 & 0.419 \\
 $\theta_2^\mathrm{polar}$ & 0.975 & 0.071 & 0.827 & 0.895 \\
 \hline
 p-values of p-values & 0.242 & 0.069 & 0.938 & 0.310 \\
 \hline
 \end{tabular}
 \caption{P-values obtained for each parameter when recovered from the actual detector noise at the time of the respective events. The last column corresponds to results from the injections used in the GW150914-like run shifted to the coalescence times used in the GW151226-like injections. We perform a KS-test comparing the set of p-values obtained for each event to a uniform distribution and report the p-value of p-values in the last row.}
  \label{real_table}
 \end{center}
\end{table*}

\captionsetup[subfigure]{labelformat=brace}

\end{document}